\documentclass[10pt,journal,compsoc,a4paper]{IEEEtran}

\usepackage{amsfonts,amstext,amsthm,amsmath,amssymb,siunitx}
\usepackage[utf8]{inputenc}
\usepackage[T1]{fontenc}
\usepackage[usenames, x11names]{xcolor}
\usepackage{titletoc} 
\usepackage[hyperindex=true, colorlinks =true, linkcolor = blue]{hyperref}
\usepackage[overload]{empheq} 
\usepackage{tikz}
\usepackage{nicefrac}

\usepackage{layouts} 

\title{Analog Computing for Molecular Dynamics}
\author{S. K\"oppel, A. Krause, B. Ulmann} 

\renewcommand{\d}{\mathrm{d}}

\newcommand{\ie}{\text{i.\,e.}\,}
\newcommand{\cf}{\text{c.\,f.}\,}

\begin{document}

\author{Sven~K\"oppel,~
	Alexandra~Krause,~
	Bernd~Ulmann~
	\IEEEcompsocitemizethanks{\IEEEcompsocthanksitem 
		The authors are with Anabrid GmbH, Am Stadtpark 3, 12345 Berlin, Germany.
		E-mail: see \url{https://www.anabrid.com}
		\IEEEcompsocthanksitem A. Krause is with FU Berlin, Arnimallee 14, 14195 Berlin.
		\IEEEcompsocthanksitem B. Ulmann is a professor at the FOM University 
         of Applied Sciences for Economics and Management. He is a guest 
         professor at the Institute of Medical Systems Biology at Ulm 
         University.
    }
	\thanks{Manuscript received March 04, 2021; revised March 04, 2021.}}

%
%

\markboth{Journal of \LaTeX\ Class Files,~Vol.~xx, No.~xx, August~2021}%
{Shell \MakeLowercase{\textit{et al.}}: Bare Advanced Demo of IEEEtran.cls for IEEE Computer Society Journals}
%



\IEEEtitleabstractindextext{%
	\begin{abstract}
	Modern analog computers are ideally suited to solving large systems of ordinary
	differential equations at high speed with low energy consumtion and limited 
    accuracy. In this article, we survey $N$-body physics, applied to a simple 
    water model inspired by force fields which are popular in molecular 
    dynamics. We demonstrate a setup which simulate a single water molecule in 
    time. To the best of our knowledge such a simulation has never been done on 
    analog computers before. Important implementation aspects of the model,
	such as scaling, data range and circuit design, are highlighted.
	We also analyze the performance and compare the solution with a numerical
	approach.
	\end{abstract}
	
	}

\maketitle

\IEEEdisplaynontitleabstractindextext

%
\IEEEpeerreviewmaketitle

\section{Introduction}

%
%
%
%
\IEEEPARstart{G}{eneral purpose analog computers} are highly efficient 
computers with respect to energy consumption and time to solution,
particularly suitable for solving
problems formualted as differential equations, albeit with limited accuracy.
Analog computers were popular
up to the 1980s and are experiencing a comeback as vital parts of hybrid computers,
which consist of a digital  and an analog computer.
The analog computer in such a setup acts as a co-processor,
speeding up simulations substantially.
The digital computer can reconfigure and parameterize the analog computer.

For a given problem, analog computers exhibit \emph{orthogonal} features in 
time vs.  power consumption: While purely analog computers (\ie not coupled with
a digital computer) cannot trade complexity vs. time as their digital 
counterparts can do, an analog computer allows power to be traded
against time~\cite{kht2020}. Given a certain power budget, 
a hybrid computer can be built which does not suffer from poor parallelization
-- one of 
the main practical drawbacks of traditional high performance digital computing 
centers. This is of particular interest for large scale scientific and 
industrial applications~\cite{Rapaport2004}
such as protein folding, a branch of research which became 
wildly popular well beyond biochemistry due to the Covid-19 race of vaccacine 
engineering and where molecular dynamics is a major theory applied
\cite{Guvench2008,Lopes2014,Mackerell2004}.

This paper is structured as follows: Section \ref{sec:water-theories} introduces an 
water modeling theories and describes our model. 
In Section~\ref{sec:analog-implementation}, we derive and demonstrate our 
analog computer implementation of this model. Section~\ref{sec:scaling} 
provides an in-depth discussion about scaling and the validity of the analog 
computer value representation.  Section~\ref{sec:benchmarks} provides an 
evaluation of the analog computer performance in terms of energy and solution
time as well as a comparison with numerical approaches. The paper closes with a 
conclusion in Section~\ref{sec:conclusion}.

\section{An $N$-body model for water}\label{sec:water-theories}
Molecular dynamics is a popular approach in chemistry, biophysics and life-sciences for
describing the dynamics of atoms and molecules. Being a classical $N$-body particle
simulation with effective physical forces between the constituent atoms (or molecules,
respectively), it has traditionally been solved numerically, using digital
computers. The first such simulations date back to the 1950s
(see for instance the reviews \cite{rahmann71,hadley12}).

\begin{figure}
	\centering
	\includegraphics[width=0.7\linewidth]{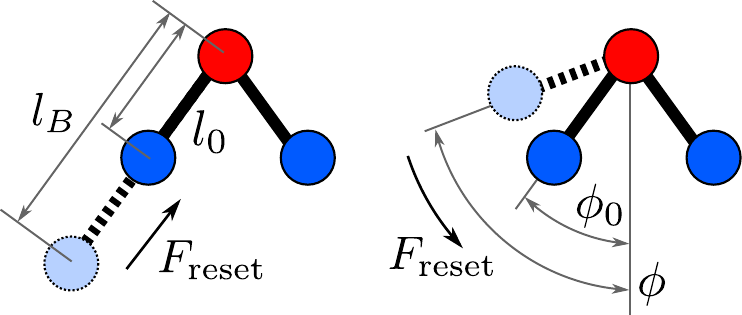}
	\caption{Intramolecular degrees of freedom in the water molecule
		modeled as harmonic displacements:
		Bond stretching (Eq.~\ref{eq:bond-stretching}) and angle
		bending (Eq.~\ref{eq:angle-bending}) between an oxygen atom (red)
		and an associated hydrogen atom (blue).
	}\label{fig:water-freedom}
\end{figure}

Mathematically, $N$-body physics is described by a set of coupled ordinary differential equations
(ODEs),
\begin{equation}\label{eq:nbody}
m_i \frac{\d^2 \vec{r}_i}{\d t^2} = - \vec \nabla_i V(\vec r_0, \dots \vec r_N) = \vec F_i(\vec r_0, \dots \vec r_N)
\,,
\end{equation}
for $N$ distinct particles with positions $\vec r_i$, masses $m_i$ and
conservative forces $\vec F_i$ generated by the potential energy $V$ with
$\nabla_i^d = \partial / \partial r^d_i$ being the $d$th component of the gradient.
The system \eqref{eq:nbody} is an ODE of second order in time and first order in space.

\subsection{Physical content: CHARMM force fields}\label{sec:charmm}
The physical content encoded in the potential $V$ is a macroscopic theory
where no quantum effects are involved. Water models  are among the most popular parts of
theoretical biophysics, as water is a very important solvent and a rather complex system.
Different water
models for digital computers have been suggested, implementing different properties that
have been experimentally studied in real water. 
These water models vary in their computational complexity and accuracy and need to be chosen
for the specific purpose of a study.
Typically, the potential is a sum of bounded (intramolecular) and
unbounded (intermolecular) contributions. Here, the typical constituents for
the CHARMM (short for \emph{Chemistry at HARvard Macromolecular Mechanics}) 
force fields will be reviewed \cite{brooks2009,brooks1983}.

First, for intramolecular contributions, one typically considers \emph{bond stretching} with
a quadratic potential (\cf Figure~\ref{fig:water-freedom})
\begin{equation}\label{eq:bond-stretching}
	V_b = \sum_{bonds} k_{b}/2~(l_b - l_0)^2
\end{equation}
with force constant $k_b$ and
bond length difference $l_b - l_0$ indicating deflection from equilibrium.
\emph{Angle bending} also models the bond angle deflection as a quadratic potential
\begin{equation}\label{eq:angle-bending}
	V_{\phi} = \sum_{angles} k_{\phi}/2~(\phi - \phi_0)^2 \,.
\end{equation}
Other contributions are obtained from
\emph{dihedral angles}, which are clockwise angles between two planes, where the planes are
spanned between three respective atoms. The potential terms include quadratic
contributions for proper and improper dihedrals. Since they only apply for molecules with
$N \geq 4$ atoms, we won't consider them here.

Next, for intermolecular contributions, one typically considers {Van-der-Waals forces},
which are modeled by Lennard-Jones potentials
\begin{equation} \label{eq:lennardjones}
V_{vdW} = \sum_{i,j | i < j} 4 \varepsilon_{i,j} \left[\left(\frac{\sigma_{i,j}}{r_{i,j}}\right)^{12} - \left(\frac{\sigma_{i,j}}{r_{i,j}}\right)^6\right]
\,.
\end{equation}
The polynomial potential \eqref{eq:lennardjones} for uncharged atoms or molecules combines
attracting and repelling forces. Indices $i$ and $j$ count over the atoms in the sum, $r_{i,j}$ is
defined as the modulus of the position vectors $r_i$ and $r_j$ for atoms $i$ and $j$.
$\varepsilon_{i,j}$ denotes the $y$-coordinate of the minimum in the potential and $\sigma_{i,j}$ describes
the $x$-coordinate of the position where $V_{vdW}=0$.

Both inter- and intramolecular electrostatic interactions of the charged nuclei
are modeled in a straightforward fashion with the Coulomb potential
\begin{equation}
  V_{el} = \sum_{i,j | i < j} \frac{q_i q_j}{4 \pi \varepsilon_0 r_{i,j}},
\end{equation}
which describes the interaction between atoms $i$ and $j$ with charges $q_i$ and $q_j$ and
their position vector difference $r_{i,j}$ as in the Lennard-Jones potential.
$\varepsilon_0$ denotes the dielectric constant.

Note that the older and currently most widely used models are rigid models which do not
allow for polarization of the water molecule. All internal degrees of
freedom are fixed and there are no dummy charges which would allow for polarization.
They differ in their parameters for the force field and are purposely fitted to
model-specific physical properties of water such as the diffusion coefficient, viscosity 
and others. Prominent examples here are the TIP3P and SPC/E water models \cite{Joergsen1983}.

Newer generations of water models introduce dummy charges to allow for a change
in polarization and dipole moment within the water molecule. All internal
degrees of freedom are still held static but the introduction of the dummy charges
allows for the negative charges to be not entirely focused on the oxygen atom but
to be located on the oxygen atom and a dummy charge below it (TIP4P) or dummy charges
simulating the valence electron pairs of the oxygen atom (TIP5P). These approaches
are computationally less costly than the introduction of flexible bounds but are 
still not used extensively as their computational costs are much higher than for
models without polarization.
%
%
%

\subsection{A single molecule demonstrator model}
We want to simulate a \emph{single} H$_2$O molecule
with internal dynamics described by Coulomb forces, bond stretching and
angle bending. Due to the limited size of the analog computer available, 
the problem is reduced to $d=2$ spatial dimensions instead of a full $d=3$ model.
Furthermore, the angle bending is modelled using a \emph{small-angle approximation},
\ie replaced by a spring between the two hydrogen atoms, which is valid
for small angular excitations. 

\begin{figure}[h]
	\centering
	\includegraphics[width=0.75\linewidth]{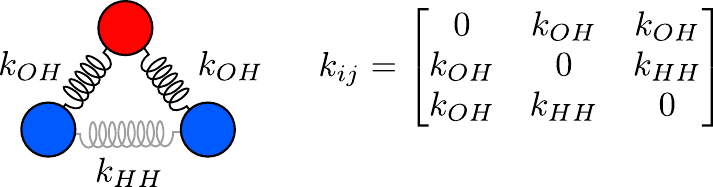}
	\caption{
		Simplified intermolecular forces modeling attracting ($k_{HH}$) and
		repelling ($k_{OH}$) forces on a purely harmonic approximation.
		This is implemented in the equations of motions \eqref{eq:eom}.
	}\label{fig:water-springs}
\end{figure}

For a single molecule in two dimensions, 
we have $N=3$ particles ($H_1, H_2, O$) and $D=2$ dimensions ($x,y$).
Generally, the indices are given by $i\in(0,1,2)\equiv(O, H_1, H_2)$
and $d\in(0,1)\equiv(x,y)$.
The equations of motions for the miniature force field are
\begin{subequations}\label{eq:eom}
\begin{align}
	\frac{\mathrm{d} \vec v_i}{\mathrm{d} t}
	&=
	\sum_{j=0}^{N-1}
	\left(
	\left(1 - \frac{l_{ij}}{r_{ij}}\right) k_{ij}
	+
	\frac{q_i q_j}{r_{ij}^3}
	\right)
	\frac{\vec r_i - \vec r_j}{m_i} \,,
	\\
	\frac{\mathrm{d} \vec r_i}{\mathrm{d} t}
	&= \vec v_i \,.
\end{align}
The spring rate matrix $k_{ij}$
\end{subequations}
models that either bond stretching ($k_{OH}$) happens between an oxygen and an hydrogen
atom or angle bending ($k_{HH}$) is taking place between the two hydrogen atoms, \cf
figure \ref{fig:water-springs}.
The rest positions $l_{ij}$ follow a similar definition with bond
length $l_{OH}$ and the HH distance $l_{HH}$.


\subsection{Physical units}\label{sec:physical-units}
We have chosen a suitable natural unit system where all constants are within
a reasonable size between zero and one. We measure the electric charges
in multiples of the elementary chage $e=1$ and thus have $q_O = q_0 =0.417$
and $q_H=q_{1,2}=-0.834$. The masses are measured in units of atomic
mass units, $m_O=m_0 = 16$ and $m_H = m_1 = m_2 = 1$. The vacuum permittivity
is set to $4\pi\varepsilon_0 = 1$. Lengths are measured in \AA, and therefore
the distance at rest (bound length) between OH is given by $l_B=0.9572 := l_{OH}$. The opening
angle at the Oxygen atom is $104^\circ=0.58\pi$, which determines the
distance at rest between HH as $l_{HH} = 2\, l_{OH} \sin(0.58\pi /2)$.

\subsection{Properties and limitations}
The \emph{characteristic frequencies} or \emph{normal modes} of
molecules~\cite{ideasOfQuantumChemistryBook} are often taken into account when reasoning
about the validity of a force field or its numerical simulation. In $d=3$
spatial dimensions, a single water molecule has $9$ well-known modes, such as
the symmetric, anti-symmetric and bending mode. Due to the
non-orthogonal geometry of our spring model \eqref{eq:eom}, one cannot expect
these modes to appear when $k_{HH}\neq 0$. This is mainly because the small
angle approximation is violated by the HH spring. On the other hand, with
$k_{HH}=0$, nothing prevents the $OH$ binding from rotating around the oxygen atom.
The angle bending approximation is a computationally cheap choice for avoding
these non-physical oscillations.

For solvent models, a number of statistically or thermodynamically derived quantities
are typically considered in benchmarks \cite{Mark2001}, such as the temperature,
potential energy and equilibirium stability of a large molecular ensemble. The
self-diffusion coefficient and the radial distribution functions for water can be
tested and are meaningful properties of a physical model (force field) and the applied
computational methods which have been used. With a single molecule, there is no meaningful equivalent.

\section{Analog computer implementation}\label{sec:analog-implementation}
The classical equations of motions \eqref{eq:nbody} are coupled ordinary differential
equations. In contrast to classical fluid dynamics, where the time evolution of
some field $F(\vec x,t)$ is determined by a partial differential equation,
ordinary differential equations of distinct particle trajectories $\vec r(t)$
can be directly mapped to electrical circuits. The traditional direct approach
represents all vectorial components $i$ of the involved particles $j$
by one connection between computing elements per {degree of freedom} $\pmb r^i_j(t)$. Following this approach, 
all intermediate computing results are also represented this way, such as velocities
$\pmb v^i_j(t)$ and any algebraic (sub-)expression encountered in the particular
analytically defined equation of motion. Note that time is kept continous while
the discretization follows the concepts of discrete and distinguishable particles in 
classical physics.

\begin{figure}[t]
	\includegraphics[width=\linewidth]{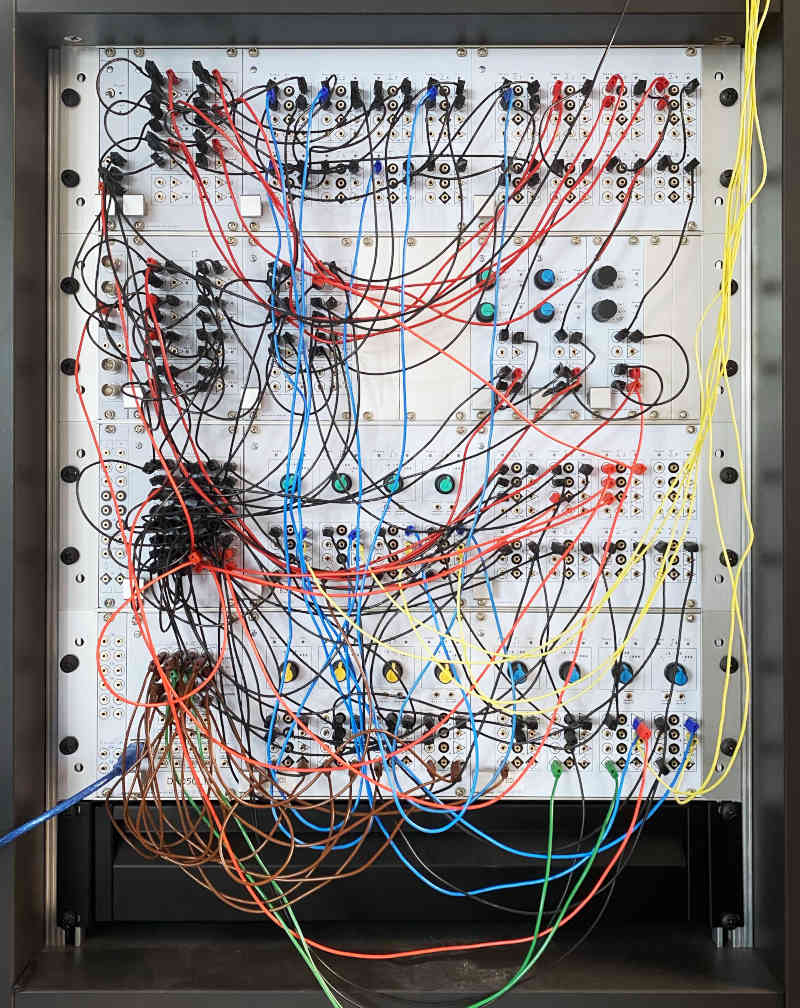}
	\caption{Photography of the front panel of a large analog computer of
		type \emph{Analog Paradigm Model-1}, implementing equations \eqref{eq:analog-equations}
		in a scaled manner (\cf section \ref{sec:analog-implementation}).
	}\label{fig:setup-photo}
\end{figure}

\subsection{Naming all constitutional parts of the equation}\label{sec:naming}
In order to implement the equation of motions on the analog
computer, one has to decompose the equation into the primitive analog computing
elements.  In general, upper indices will indicate the vector index
$\cdot^d$, while lower indices $\cdot_i$ indicate the particle index.
Bold face is used to separate the name of a variable (connection) whereas indices
are given in regular form, which is particular relevant for the multiplicative
inverses, such as the distance matrix $\pmb{r}^{\pmb{-1}}_{ij}$. It is
computed as following:
\begin{subequations}\label{eq:analog-equations}
	\begin{equation}\label{eq:rst}
		\pmb r_{ij}^d = r_i^d - r_j^d
		\,,\quad
		\pmb{s}^d_{ij} = \left( \pmb r_{ij}^d \right)^2
		\,~~\text{and}~~
	\end{equation}
	\begin{equation}
		\pmb{t}_{ij} = \textstyle \sum_k^D \pmb{s}_{ij}^k
		\,,~~\text{so that}~~
		\pmb{r}^{\pmb{-1}}_{ij} = 1 / \sqrt{\pmb t_{ij}}
		\,.
	\end{equation}
	Note that the coordinate difference components $\pmb r^d_{ij} = -\pmb r^d_{ji}$ are
	antisymmetric. The distance matrix $\pmb{r}_{ij}$ and its multiplicative inverse
	$\pmb{r}^{\pmb{-1}}_{ij}$ are symmetric ($\pmb{r}_{ij} = \pmb r_{ji}$) and
	traceless ($\pmb r_{ii}=0$). Thus one only has to compute 
	$\pmb{r}_{ij} \in \mathbb{R}^{M \times D}$ many elements, with $M=(N^2 - N)/2 = 3$.
	The actual force components 
	\begin{equation}\label{eq:physical-f}
		\pmb{F}_{ij}^d = \left(\left[l_{ij} {\pmb r}_{ij}^{\pmb{-1}} - 1 \right] k_{ij}   +  \frac{q_i q_j}{\varepsilon_0} ({\pmb r}_{ij}^{\pmb{-1}})^{3} \right) \pmb r_{ij}^d
	\end{equation}
	are antisymmetric and traceless
		($\pmb F^d_{ij} = - \pmb F^d_{ji}$ and
		$\pmb F^d_{ii} = 0$ for all $i,j,d$).
	Also note that the bond stretching is only
	applied between H and O atoms of a single molecule, which is encoded in
	$\Delta_{ij}$.
	The velocity and position components can be determined by analog time integration
	\begin{equation}\label{eq:physical-vr}
		\pmb v_i^d
		= -\int_t \mathrm d t \sum_j^N \left(- \frac {\pmb {F}_{ij}^d} {m_i}\right)
		,\quad
		\pmb r_i^d
		=-\int_t \mathrm d t ~(-\pmb v_i^d),
	\end{equation}
	where the notation is oriented on the negating nature of the analog computing 
    elements.\footnote{Analog computer modules such as summers and integrators perform
    an implicit change of sign which is due to the actual technical implementation of 
    these computing elements.}
\end{subequations}

\subsection{Machine realization}
The set of equations \eqref{eq:analog-equations} is directly mapped to the general purpose
analog computing elements of the \emph{Analog Paradigm Model-1} computer we used in this simulation
\cite{Model1Handbook}. The overall circuit
is given in Appendix~\ref{apx:circuit} and is a 1:1 translation of the mathematical expressions
\eqref{eq:analog-equations}. We would like to emphasize that on this level of abstraction,
the abstract syntax tree of the mathematical expressions is \emph{identical} to the equivalent circuit.
The physical realization of this particular analog computer program is shown in figure~\ref{fig:setup-photo}. The
machine is equipped with a \emph{Hybrid Controller}
(it can be identified by its USB cable in the lower left of figure~\ref{fig:setup-photo}).
The system can be fully configured from an attached digital computer
to set the model's constants and parameterers; there are more then
30 digital potentiometers involved, each with a resolution of 10 bit.

\section{Circuit scaling}\label{sec:scaling}
The question of value ranges and domain sizes is crucial in any (analog or digital) computer
simulation of a physical problem. This topic is typically not given much attention in presentations
of novel numerical codes. When implementing analog circuits, ensuring the correct scaling of
the solution is an inherent part of the problem mapping. Various aspects of
this scaling process are discussed here.

\subsection{Analog number representation}\label{sec:intro-scaling}
For number representation on computers, it is essential that quantities do not
exceed the valid range of values. For instance, a signed 32 bit integer
can store values between 0 and $2^{31} \approx 10^{10}$. In contrast, the IEEE 754
floating point number representation allows to represent numbers up to
$2^{127} \approx 10^{38}$, which comes with a loss of precision.
When doing numerical computations, any number outside the valid data range results in an
overflow and an invalid result. However, on a digital computer, there is always
the possibility to use longer bit sequences to represent numbers. For instance, one could
use a double-precision floating point number representation with a width of 64 bit,
allowing numbers up to $2^{1023} \approx 10^{308}$. The same applies
for numbers which are bounded by nature, for instance percentages or saturations.
If a number $p \in [0,1]$ is to be represented on a digital computer, then the 32 bit
signed integer will naturally represent $p$ with a uniform accuracy of
$\Delta p \approx 10^{-38}$. In contrast, 32 bit single-precision floating point has
a precision $\Delta p \approx 10^{-20}$ for values down to $10^{-12}$ and up to
$\Delta p \approx 10^{-8}$ for values close to $1$.

The success and widespread use of floating point numbers makes number representation a
topic, which numerical scientists only have to deal in special cases, for instance when
it comes to instabilities, catastrophic cancellation or the like. In day-to-day work, the occurence
of an out-of-bound value (represented as \emph{NaN}, not a number, in the IEE754
standard) is typically a sign of a programming error, such as computing $1/0$ resulting from
an uninitialized or in some way ``illegal'' denominator.

In electronic analog computers, numbers are typically represented as voltages or
currents. Given the fixed physical structure of such a computer, these quantities
are bounded in principle. On an information theoretical level,
such computing elements are called ``bounded in, bounded out'' (BIBO) \cite{Westphal1995}. The
illegality of non-bounded values is similar as to the behaviour of number representations in
digital computing. For the time being, let's assume the valid range for a value in 
an analog computer to be $[-1,+1]$.

Analog computers have an inherent limited precision of
about three to four decimal places
\cite{Sarpeshkar1998}. In the above convention, this yields a precision $\Delta p \approx 10^{-4}$ which
is roughly the resolution of a half-precision floating point number or 10 to 16 bits of information.

In order to answer the question of whether simulating water molecules on an analog computer can compare
to current digital simulations, given the limitations in resolution of the analog computer,
we assume the Coulomb potential terms to be the limiting factor. Keep in mind that in the present
study we can neglect the Lennard-Jones potential as we are only modelling a single molecule.

\subsection{Coordinate units}
To improve readability, we assume a physical square simulation domain
of size $\Omega := [-L,+L]^D$ with some dimensionful length scale $L$ in $D$ spatial dimensions with
any physical coordinate $\pmb r^d_i \in \Omega$. Scaled and dimensionless units can be easily
recovered as $\hat r := r / L \in \hat \Omega$ contained in a unit domain $\hat \Omega = [-1,+1]^D$.
In practice that means that initial values have to be scaled according to $\hat r_i^d(0) = r_i^d(0) / L$ and
$\hat v^i_d(0) = v_i^d(0) / L$. Also the distances at rest
$l_{OH}$, $l_{HH}$ and the de facto Coulomb coupling constant $\varepsilon_0$ have to be
chosen in physical units before transformation.

Assuming that each mapping in \eqref{eq:rst} has a bounded input on the right hand side
of the definition, the following scaling has to be applied to ensure bounded output:
%
%
\begin{equation}\label{eq:rst-scaled}
	\hat{\pmb r}_{ij}^d := \frac{\hat r_i^d - \hat r_j^d}{\alpha}
	\,,\quad
	\hat{\pmb{s}}^d_{ij} := \left( \hat {\pmb r}_{ij}^d \right)^2
	\,,\quad
\end{equation}
\begin{equation}
	\hat{\pmb{t}}_{ij} := \frac{\textstyle \sum_k^D \hat{\pmb{s}}_{ij}^k}{\delta^2}
	\,,\quad
	\hat{\pmb{r}}^{\pmb{-1}}_{ij} := 1 / \left({\beta\sqrt{\hat{\pmb t}_{ij}}}\right)
	\,.
\end{equation}
Three factors were introduced:
Two geometric ones ($\alpha$ and $\delta$) resulting from the dimensionless coordinate representation 
and an internal one ($\beta$) to deal with the multiplicative inverse (reciprocal).
These scaling factors sum up to
\begin{equation}
	\hat{\pmb {r}}^{\pmb{-1}}_{ij} := \gamma\, \pmb {r}^{\pmb{-1}}_{ij}
\end{equation}
	with overall scaling factor $\gamma :=\alpha\delta / \beta$.
Note that applying this scaling means that $\hat{\pmb r}^{-1}_{ij} \neq 1/|\hat{ \pmb{r}}_i - \hat{\pmb r}_j|$, \ie
a clear relationship between the ``hatted'' quantities $\hat{\pmb r}_{\dots}$ no longer exists.
The neccessity and implications of the individual factors are discussed below.

\subsubsection{Geometric scaling}
The factor $\alpha$ has two contributions determined by the maximum possible distance of
two particles in both $D$ dimensions ($d_\text{max} = \sqrt{D}$) and one dimension, where
$d_\text{max} = 2$ and $1 \leq \alpha \leq 2$ is a sensible choice.
The factor $\delta$ results from the sum of $D$ scaled quantities. A useful choice might be
$\delta = \sqrt{D}$. Both factors $\alpha$ and $\delta$ are \emph{optional} as
they might be applied but their underlying scaling problems can also be neglected in practice
if no overflow problems occur. In order to simplify the reasoning, the optional factors are neglected
for the rest of the text, so $\alpha=\delta=1$ and thus $\gamma=1/\beta$.

\subsubsection{Inverse-square law scaling}
The factor $\beta$ is due to the necessity of computing multiplicative inverses on BIBO unit
domains. It is crucial and must not be omitted.
Displaying an inverse-square law such as the Coulomb potential
$V_{ij}(d) \sim 1/r_{ij}$ or the electrostatic force $F_{ij}(d) \sim 1/r_{ij}^2$ with
$r_{ij} = |\vec r_i - \vec r_j|$ being the distance between two particles $i$ and $j$
on an analog computers is challenging, since it is
immediately $1/r_{ij} > 1$ for $r_{ij}<1$. 
The approach in \eqref{eq:rst-scaled} is to apply a scaling factor $1/\beta \ll 1$ on
$\pmb {r^{-1}}$. 
This allows the distance component $\pmb t_{ij}$ to become as small as $\beta^{-2}$
(because for smaller $\pmb t_{ij}<\beta^{-2}$, again $\pmb{r^{-1}}>1$). Naturally,
this introduces an \emph{ultraviolet cutoff} at $\beta^{-1}$. If two particles
come closer than $\beta^{-1}$, their distances can no longer be represented on the
computer. Cutoffs are a well-understood phenomenon in theoretical physics and also occur
in numerical simulations.

\subsection{In-circuit scaling}
The scaling coefficients $\alpha, \beta, \delta$ are not relevant outside the
computation. Instead, they only affect the potentiometer values. By convention,
the scaling factors (lowercase greek letters as $\alpha$) are typically outside the BIBO range
$[-1,+1]$, while the potentiometer values (uppercase Fraktur letters as $\mathfrak A$) are within.

\subsubsection{Force scaling}
The system force \eqref{eq:physical-f} is scaled as
\begin{equation}
\hat{\pmb F}^d_{ij} := \zeta \frac{{\pmb F}^d_{ij}}{\beta^3} 
\end{equation}
The factor $\beta^3$ results from the linear dependency on 
$\pmb r^{-3}$, which implies carrying on the scaling. The factor
$\zeta>1$ is an non-nphysical scaling factor introduced in order to allow force amplification
in the subsequent steps.

By replacing unscaled quantities in \eqref{eq:physical-f} by their scaled counterparts
of the previous section, the potentiometer values are determined as:
\begin{equation}
\hat{\pmb F}^d_{ij} := \Bigg[
\left(\frac{l_{ij}}{\beta^2}~\frac{\hat{\pmb{r}}^{\pmb{-1}}_{ij}}{\beta}
- \frac{1}{\beta^3}\right) k_{ij}  
+ \frac{q_i q_j}{\varepsilon_0}  \frac{\hat{\pmb r}^{\pmb{-3}}_{ij}}{\beta^3}
\Bigg]  \zeta~ {\hat {\pmb r}^d_{ij}}
\end{equation}
With the intention to define parameters $\frak{P}$, which directly map on a single
potentiometer with range $0 \leq \frak{P} \leq 1$ on the \emph{Model-1}, we introduce
the quantities
\begin{equation}
	\mathfrak{R}_{ij} = \zeta \frac{l_{ij} k_{ij}}{\beta^2} 
	\,,\quad
	\mathfrak{U}_{ij} =  \zeta \frac{k_{ij}}{\beta^3} 
	\,,\quad
	\mathfrak{C}_{ij} = \zeta \frac{|q_i q_j|}{\varepsilon_0} \,.
\end{equation}
The scaled force is then 
\begin{equation}\label{eq:force-scaling}
	\hat{\pmb F}^d_{ij} =
	\left(
	{\mathfrak{R}_{ij}}~\frac{\hat{\pmb{r}}^{\pmb{-1}}_{ij}}{\beta}
	- {\mathfrak{U}_{ij}} 
	\pm \mathfrak{C}_{ij}\,  \frac{\hat{\pmb{r}}^{\pmb{-3}}_{ij}}{\beta^3}
	\right) {\hat{\pmb r}^d_{ij}}  \,.
\end{equation}
One has to 
take care of the sign in front of $\mathfrak C_{ij}$, which is $-1$ for OH interactions and $+1$ for HH
interactions. The required negation is indicated by $\pm$ in \eqref{eq:force-scaling}. 
When following the scaling argument $\delta$ of the previous section, \eqref{eq:force-scaling} 
would also require a factor $1/3$ to compensate the sum of three scaled quantities.
However, due to the different orders of magnitude of the coupling constants, this
scaling choice was not considered here.

\subsubsection{Time integration}
The standard analog time integration \cite{ap2} reads 
\begin{equation}\label{eq:model1-integration}
	e_0 = - \left( \int_0^t k_0 \mathrm d t' \sum_{r=1}^n a_r e_r(t')  \right) - e(0)
\end{equation}
with initial condition $e(0)$, inputs $e_r$, output $e_0$, input weights $a_r$,
time scale factor $k_0$ and machine time $t'=t/k_0$.
On the Model-1 analog computer, possible input weights are $a_i\in \{1,10\}$
and $k_0 \in \{10, 100, 1000\}$. These have to be chosen at programming time.
For the velocity and coordinate integration, we can compute physical quantities
$\hat{\pmb v}_i^d \equiv \pmb v_i^d$ and $\hat{\pmb r}_i^d \equiv \pmb r_i^d$ by carefully
reversing all scaling operations. We do so by casting \eqref{eq:physical-vr} as 
\begin{align}
	\pmb-\hat{\pmb v}_i^d &= -\int_0^t k_0 \, \mathrm d t 
    \sum_j^N \left(a_i \frac{1}{m_i} \, {\pmb F_{ij}^d}\right)
    \\
    &=  -\int_0^t k_0 \, \mathrm d t 
    \sum_j^N \left(a_i \frac{1}{m_i} \frac{\beta^3}{\zeta}\, {\hat{\pmb F}_{ij}^d}\right)
    \\
    &=  -\int_0^t k_0 \, \mathrm d t 
	\sum_j^N \left(\mathfrak{F}_i\, {\hat{\pmb F}_{ij}^d}\right)\,.
\end{align}
We introduced the potentiometer value
\begin{equation}
	\mathfrak F_i = \frac{1}{m_i} \frac{\beta^3}{\zeta} \frac{1}{a_i} \,.
\end{equation}
Thanks to moving the integrator weight $a_i$ into $\mathfrak F_i$, it can be ensured
that $\mathfrak F_i \leq 1$ for many weight choices. If this is not sufficient, one can
still make use of the timescale factors $k_0$ for an additional degree of freedom in scaling. 
In our setup all integrators
have the same $k_0$ which is thereby solely responsible for the speed of integration
(or simulation vs. realclock time ratio). On future chip-level generations,
$k_0 \approx 10^7$ is expected \cite{kht2020}.
The velocity integration is then trivially given by
\begin{equation}
	\hat{\pmb r}_i^d
	=-\int_0^t k_0 \mathrm d t ~(\pmb -\hat{\pmb v}_i^d) \,.
\end{equation}
%
%
%
\subsubsection{Numerical potentiometer values}
With $\beta=5$ and the physical units given in section \ref{sec:physical-units},
we obtain $\mathfrak{R}_{HH} = 0.036 \zeta$, $\mathfrak{R}_{OH} = 0.04\zeta$,
$\mathfrak{C}_{ij}=(\pm)0.0139 \zeta$ and 
$\mathfrak{U}_{ij} = 0.008\zeta$. These small
numbers suggest to upscale the forces with $\zeta=30$. This results in 
$\mathfrak{F}_0 = \beta^3/\zeta/m_O = 125 / 16 / 30 = 0.2594$ with $a_0=1$ and
$\mathfrak{F}_1 = \mathfrak{F}_2 = \beta^3 / m_H / \zeta /10 = 125/30/10  = 0.4166$ with $a_1=a_2=10$.



\subsection{Conclusions on the dynamical range}

In the last section, we came up with a value scaling, which introduces a cutoff at
a length scale $\beta^{-1}$. With typical choices $\beta \sim 10$, this introduces
basically a minimal length the order of $L_\text{min} = 1/\beta \sim 0.1$.
Equally, the maximum length in the simulation domain is determined by
the BIBO domain size, \ie $L_\text{max} = 1$.
Note that the minimal length is \emph{not} a rounding/resolution error (as a $\Delta x$ would be in a numerical approach).
The computational exactness (rounding error) of an analog computer is of the order of $10^{-5} \ll L_\text{min}$.

$L_\text{min}$ provides a minimal distance between two atoms, so it is the limiting
factor in terms of precision. 
Chemical bonds like the
O-H distance are of the order of magnitude of 1 \AA\ while the valid range of
the entire water box runs from $-1$ analog units to $+1$ analog units. Identifying
$L_\text{min}$ analog units with 1 \AA\ in the water box a maximum side length for
a water box of 20 \AA\ or 2\,nm is possible.
Most water boxes are in the nanometer range, so this value representation
might be applicable to tackle real world simulations. Nevertheless we acknowledge
that it might be too small to contain large macromolecules and different value
representations or digital assisted approaches may be neccessary in future work.

\section{Complete Circuit}\label{apx:circuit}
The overall circuit of the setup shown in
Figure \ref{fig:setup-photo} is described below.
All indices and sums are written out and are fully scaled.
Note that we moved some constants between the definitions
(for instance between $F$ and $C$). 

A triangular shape denotes a summer (with implicit change of sign), while a polygon with
a large $\Pi$ represents a multiplier, divider or square root component.
Circles represent  coefficient potentiometers and rectangles with attached triangles are integrators.

\subsection{Computing distance components $r^i_{jk}$, $s^i_{jk}$ and $t_{jk}$}
\begin{equation}\label{eq:computing_rst}
	\begin{aligned}
		r^0_{01} &= -(-r_0^0) + (-r_1^0)   \\
		r^0_{02} &= -(-r_0^0) + (-r_2^0)   \\
		r^0_{12} &= -(-r_1^0) + (-r_2^0)   \\
		r^1_{01} &= -(-r_0^1) + (-r_1^1)   \\
		r^1_{02} &= -(-r_0^1) + (-r_2^1)   \\
		r^1_{12} &= -(-r_1^1) + (-r_2^1)
	\end{aligned}
	\qquad
	\begin{aligned}
		s^0_{01} &= (r^0_{01})^2   \\
		s^0_{02} &= (r^0_{02})^2    \\
		s^0_{12} &= (r^0_{12})^2    \\
		s^1_{01} &= (r^1_{01})^2    \\
		s^1_{02} &= (r^1_{02})^2    \\
		s^1_{12} &= (r^1_{12})^2  
	\end{aligned}
\end{equation}
\begin{equation*}
	t_{01} = s^0_{01} + s^1_{01} \quad~~~ 
	t_{02} = s^0_{02} + s^1_{02} \quad~~~ 
	t_{12} = s^0_{12} + s^1_{12} 
\end{equation*}

\begin{figure}[h!]
	\center
	\includegraphics[width=0.75\linewidth]{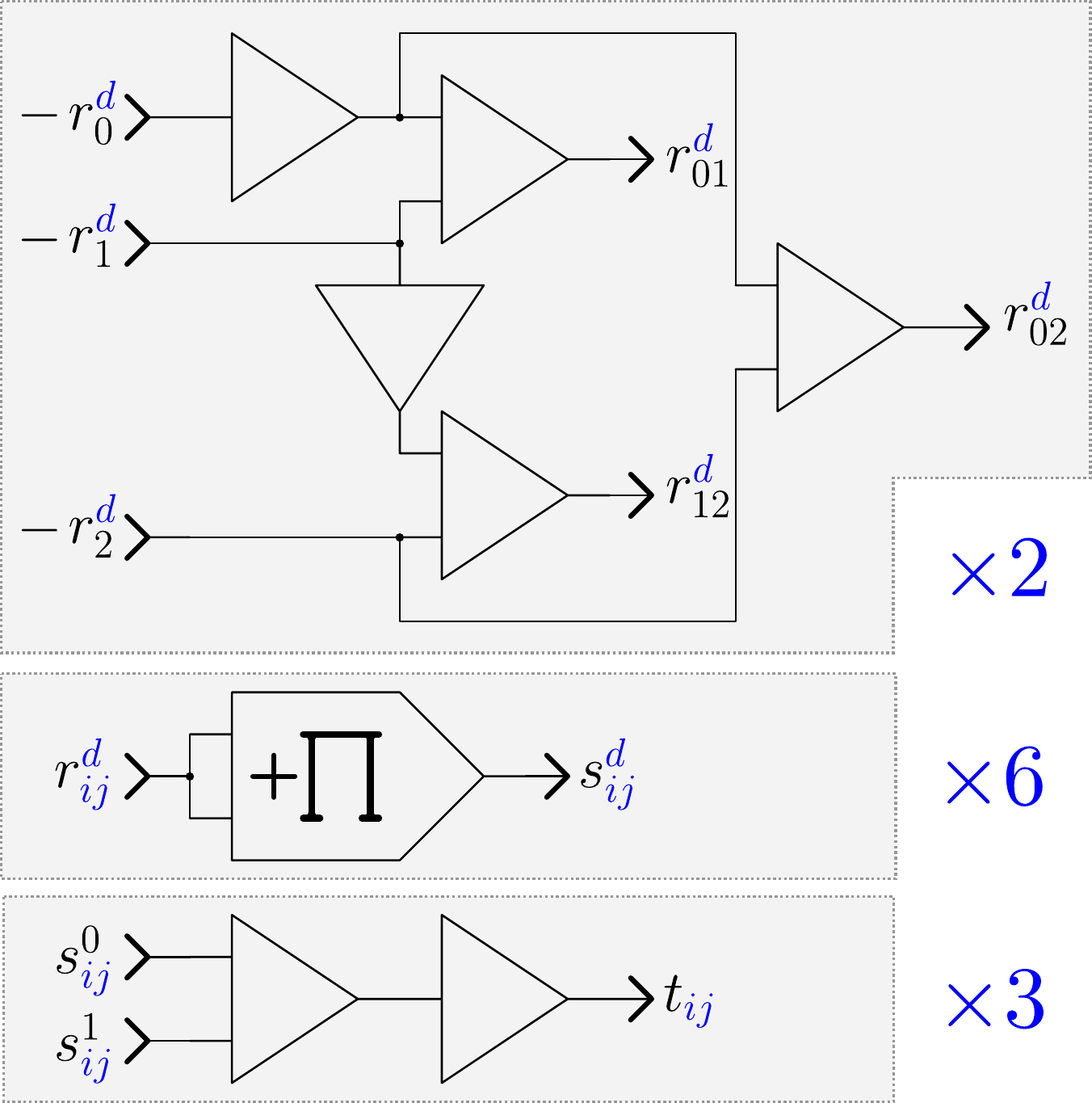}
	\caption{Computing $r^i_{jk}$, $s^i_{jk}$, and $t_{jk}$, as in
		equation \eqref{eq:computing_rst}.}
	\label{pic_h2o_setup_1}
\end{figure}

\subsection{Computing inverse distances $\pmb{r^{-1}}$ and their cubes}
\begin{figure}[h!]
	\centering
	\includegraphics[width=0.85\linewidth]{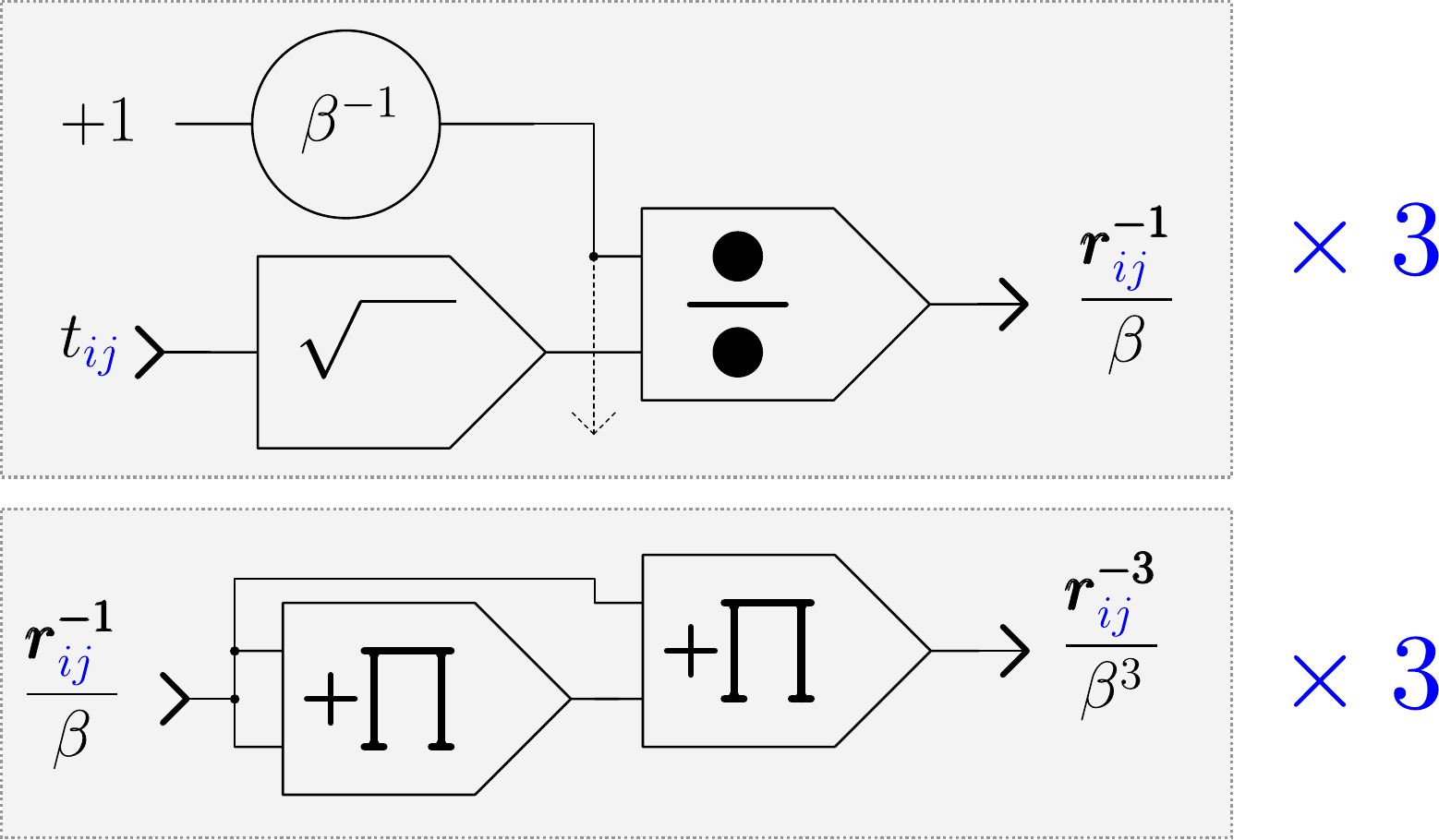}
	\caption{Computing $\frac{r^{-1}_{jk}}{\beta}$ and $\frac{r^{-3}_{jk}}{\beta^3}$,
		according to equation \eqref{eq:computing_rmc}.}
	\label{pic_h2o_setup_2}
\end{figure}

The scaling with $\beta$ is applied in the circuit (figure \ref{pic_h2o_setup_2})
\emph{before} applying the division. A single potentiometer for $\beta^{-1}$ is
sufficient as an input for all three divisions.

\begin{equation}\label{eq:computing_rmc}
	\begin{aligned}
		\frac{\pmb{r}^{\pmb{-1}}_{01}}{\beta} &= \beta^{-1}/\sqrt{t_{01}} \\
		\frac{\pmb{r}^{\pmb{-1}}_{02}}{\beta} &= \beta^{-1}/\sqrt{t_{02}} \\ 
		\frac{\pmb{r}^{\pmb{-1}}_{12}}{\beta} &= \beta^{-1}/\sqrt{t_{12}}
	\end{aligned}
	\qquad\qquad
	\begin{aligned}
		\frac{\pmb{r}^{\pmb{-3}}_{01}}{\beta^3} &= ( \pmb{r}^{\pmb{-1}}_{01} )^3 / \beta^3 \\ 
		\frac{\pmb{r}^{\pmb{-3}}_{02}}{\beta^3} &= ( \pmb{r}^{\pmb{-1}}_{02} )^3 / \beta^3 \\
		\frac{\pmb{r}^{\pmb{-3}}_{12}}{\beta^3} &= ( \pmb{r}^{\pmb{-1}}_{12} )^3 / \beta^3
	\end{aligned}
\end{equation}

\subsection{Computing the forces}
\begin{equation}\label{eq:computing_fijk}
	\begin{alignedat}{5}
		\frac{\pmb F^0_{01}}{\beta^3}
		&=
		\Big(
		{\mathfrak{R}_{01}}~\frac{\pmb{r}^{\pmb{-1}}_{01}}{\beta}
		- {\mathfrak{U}_{01}}
		~&-~ \mathfrak{C}_{01}  \frac{\pmb{r}^{\pmb{-3}}_{01}}{\beta^3}
		\Big) r^0_{01}
		\\
		\frac{\pmb F^0_{02}}{\beta^3}
		&=
		\Big(
		{\mathfrak{R}_{02}}~\frac{\pmb{r}^{\pmb{-1}}_{02}}{\beta}
		- {\mathfrak{U}_{02}}
		&-~ \mathfrak{C}_{02}  \frac{\pmb{r}^{\pmb{-3}}_{02}}{\beta^3}
		\Big) r^0_{02}
		\\
		\frac{\pmb F^0_{12}}{\beta^3}
		&=
		\Big(
		\mathfrak{R}_{12}~\frac{\pmb{r}^{\pmb{-1}}_{01}}{\beta}
		- \mathfrak{U}_{12}
		&+~ \mathfrak{C}_{12}  \frac{\pmb{r}^{\pmb{-3}}_{12}}{\beta^3}
		\Big) r^0_{12}
		\\
		\frac{\pmb F^1_{01}}{\beta^3}
		&=
		\Big(
		{\mathfrak{R}_{01}}~\frac{\pmb{r}^{\pmb{-1}}_{01}}{\beta}
		- {\mathfrak{U}_{01}}
		~&-~ \mathfrak{C}_{01}  \frac{\pmb{r}^{\pmb{-3}}_{01}}{\beta^3}
		\Big) r^1_{01}
		\\
		\frac{\pmb F^1_{02}}{\beta^3}
		&=
		\Big(
		{\mathfrak{R}_{02}}~\frac{\pmb{r}^{\pmb{-1}}_{02}}{\beta}
		- {\mathfrak{U}_{02}}
		&-~ \mathfrak{C}_{02}  \frac{\pmb{r}^{\pmb{-3}}_{02}}{\beta^3}
		\Big) r^1_{02}
		\\
		\frac{\pmb F^1_{12}}{\beta^3}
		&=
		\Big(
		{\mathfrak{R}_{12}}~\frac{\pmb{r}^{\pmb{-1}}_{02}}{\beta}
		- {\mathfrak{U}_{12}}
		&+~ \mathfrak{C}_{12}  \frac{\pmb{r}^{\pmb{-3}}_{12}}{\beta^3} \Big) r^1_{12}
	\end{alignedat}
\end{equation}

\begin{figure}[h!]
	\centering
	\includegraphics[width=\linewidth]{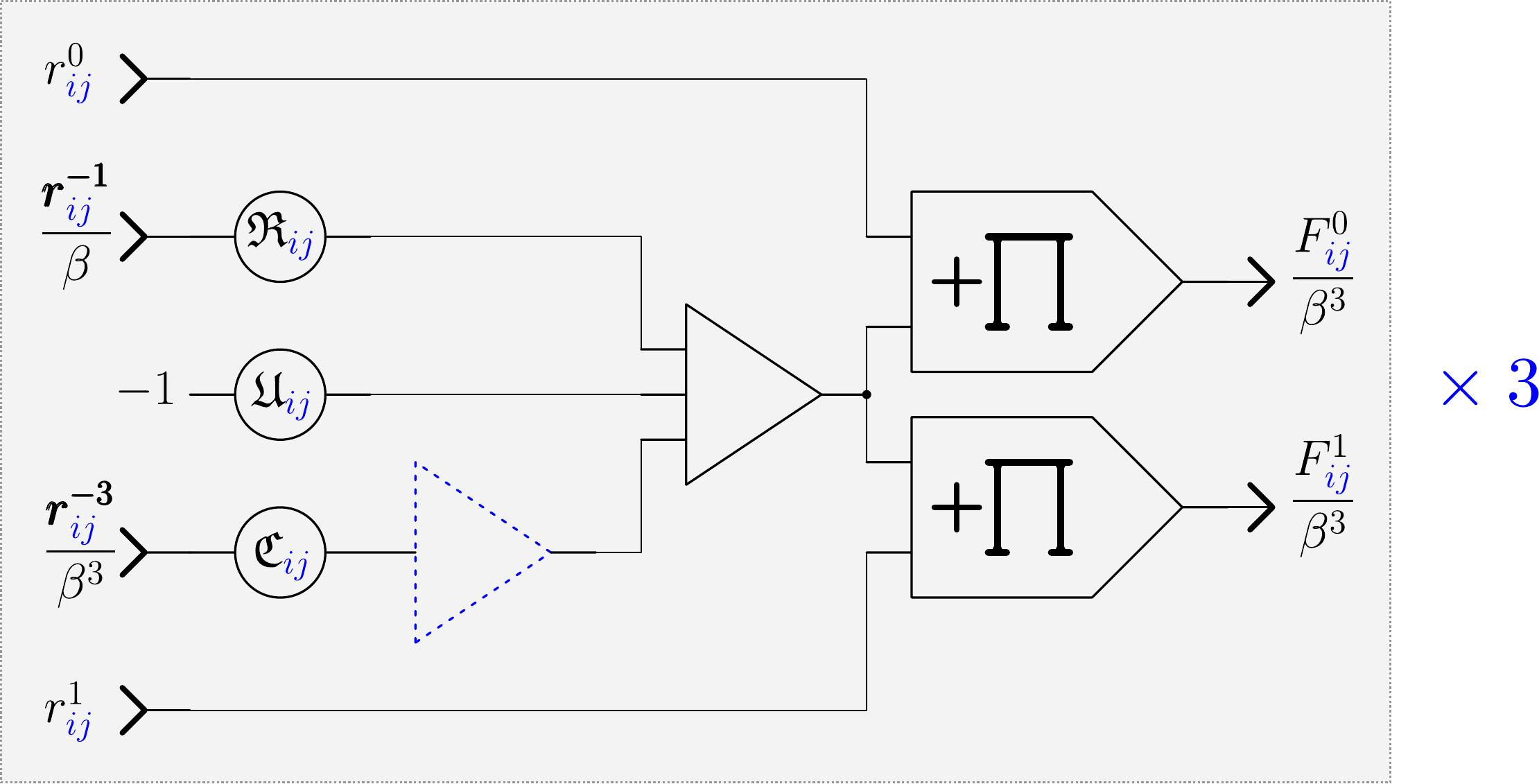}
	\caption{Computing $\frac{F^i_{jk}}{\beta^3}$ as in equation \eqref{eq:computing_fijk}.
		Note that the negator indicated in dotted lines is only placed when computing
		$F^d_{01}$ and $F^d_{02}$ but not $F^d_{12}$.}
	\label{pic_h2o_setup_3}
\end{figure}

\subsection{Computing velocities and positions}
Note how the signs in front of $\mathfrak F_i$ result from symmetries carried
out (\cf section \ref{sec:naming}).
\begin{figure}[b]
	\centering
	\includegraphics[width=0.8\linewidth]{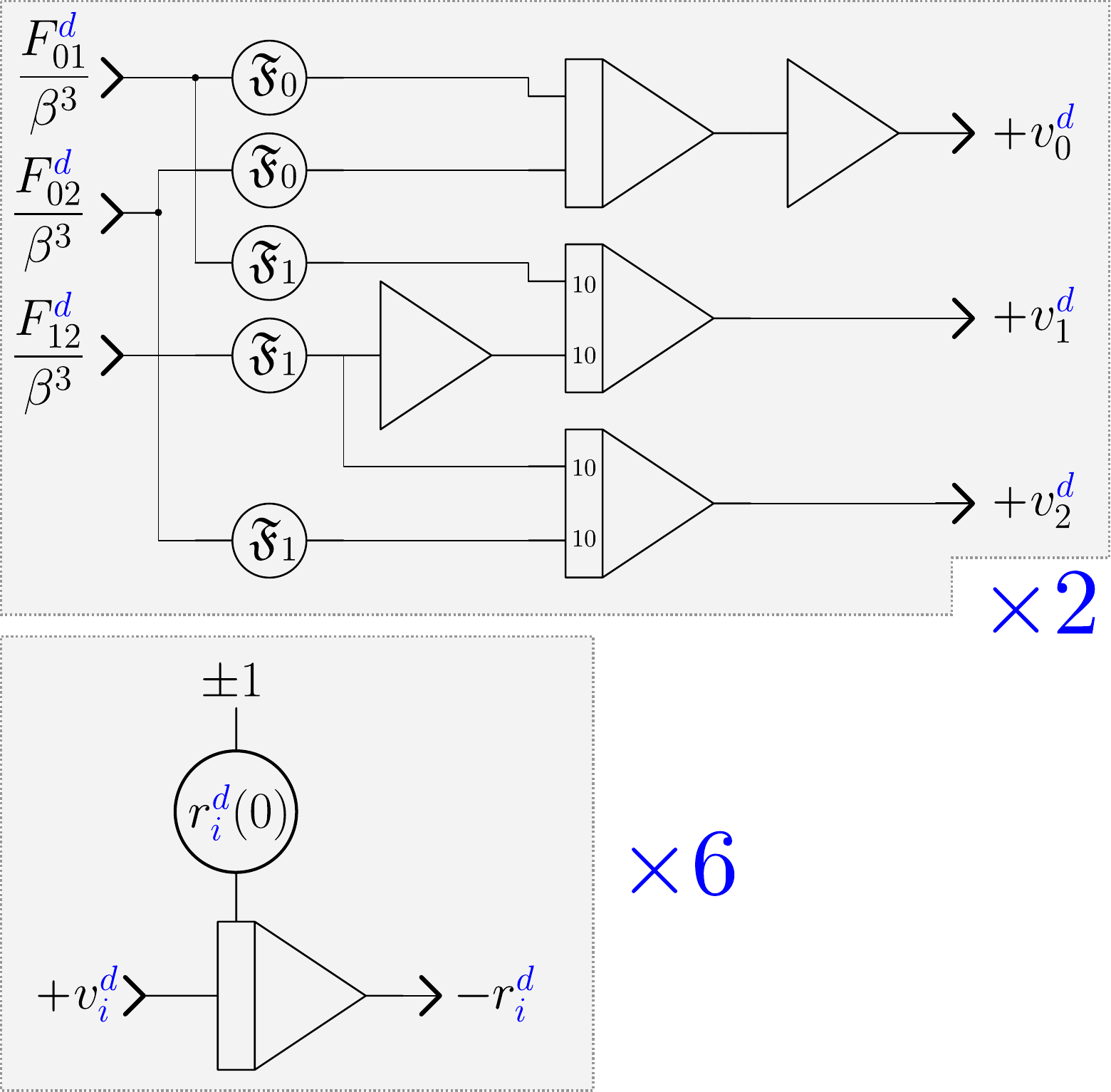}
	\caption{Computing $v^i_j$ and $r^i_j$, as in equation
		\eqref{eq:computing_r}.}
	\label{pic_h2o_setup_4}
\end{figure}
\begin{equation}\label{eq:computing_v}
	\begin{aligned}
		-\pmb v_0^0 
		&= -\int^t_0  k_0\,\mathrm d t \left( +\mathfrak{F}_0 \cdot \beta^3 ~ \frac{F_{01}^0}{\beta^3} + \mathfrak{F}_0 \cdot \beta^3 ~ \frac{F_{02}^0}{\beta^3} \right)
		\\
		-\pmb v_1^0 
		&= 
		-\int^t_0  k_0\,\mathrm d t \left( -\mathfrak{F}_1 \cdot \beta^3 ~ \frac{F_{01}^0}{\beta^3} +  \mathfrak{F}_1 \cdot \beta^3 ~ \frac{F_{12}^0}{\beta^3}  \right)
		\\
		-\pmb v_2^0 
		&=
		-\int^t_0  k_0\,\mathrm d t \left( -\mathfrak{F}_1 \cdot \beta^3 ~ \frac{F_{02}^0}{\beta^3} -  \mathfrak{F}_1 \cdot \beta^3 ~ \frac{F_{12}^0}{\beta^3}  \right)
		\\
		\Leftrightarrow 
		+\pmb v_2^0 
		&=
		+\int^t_0  k_0\,\mathrm d t \left( +\mathfrak{F}_1 \cdot \beta^3 ~ \frac{F_{02}^0}{\beta^3} +  \mathfrak{F}_1 \cdot \beta^3 ~ \frac{F_{12}^0}{\beta^3}  \right)
	\end{aligned}
\end{equation}
\begin{equation}
	\begin{aligned}
		-\pmb v_0^1
		&= -\int^t_0 k_0\,\mathrm d t \left(+ \mathfrak{F}_0 \cdot \beta^3 ~ \frac{F_{01}^1}{\beta^3} + \mathfrak{F}_0 \cdot \beta^3 ~ \frac{F_{02}^1}{\beta^3} \right) 
		\\
		-\pmb v_1^1 
		&=
		-\int^t_0 k_0\,\mathrm d t \left(- \mathfrak{F}_1 \cdot \beta^3 ~ \frac{F_{01}^1}{\beta^3} +  \mathfrak{F}_1 \cdot \beta^3 ~ \frac{F_{12}^1}{\beta^3}  \right)
		\\
		-\pmb v_2^1
		&=
		-\int^t_0  k_0\,\mathrm d t \left( -\mathfrak{F}_1 \cdot \beta^3 ~ \frac{F_{02}^1}{\beta^3} -  \mathfrak{F}_1 \cdot \beta^3 ~ \frac{F_{12}^1}{\beta^3}  \right)
		\\
		\Leftrightarrow 
		+\pmb v_2^1
		&=
		+\int^t_0  k_0\,\mathrm d t \left( +\mathfrak{F}_1 \cdot \beta^3 ~ \frac{F_{02}^1}{\beta^3} +  \mathfrak{F}_1 \cdot \beta^3 ~ \frac{F_{12}^1}{\beta^3}  \right)
	\end{aligned}
\end{equation}
In our circuit, the initial data for the velocities are always zero,
while initial data for the positions are fed in as initial conditions for
the integrators (see figure \ref{pic_h2o_setup_4}).
\begin{equation}\label{eq:computing_r}
	\begin{aligned}
		\pmb r_0^0 &= -\int^t_0 (-v_0^0) ~ k_0 \, \mathrm d t - r^0_0(0)\,, \\
		\pmb r_1^0 &= -\int^t_0 (-v_1^0) ~ k_0 \, \mathrm d t  - r^1_0(0)\,, \\
		\pmb r_2^0 &= -\int^t_0 (-v_2^0) ~ k_0 \,  \mathrm d t - r^2_0(0) \,, 
		\\
		\pmb r_0^1 &= -\int^t_0 (-v_0^1) ~ k_0 - r^0_1(0)\, \mathrm d t  \,, \\
		\pmb r_1^1 &= -\int^t_0 (-v_1^1) ~ k_0 - r^1_1(0)\,  \mathrm d t  \,, \\
		\pmb r_2^1 &= -\int^t_0 (-v_1^1) ~ k_0 - r^2_1(0)\,  \mathrm d t  \,.
	\end{aligned}
\end{equation}

\section{Benchmarking the analog implementation}\label{sec:benchmarks}
\begin{figure}[t!]
	\includegraphics[width=\linewidth]{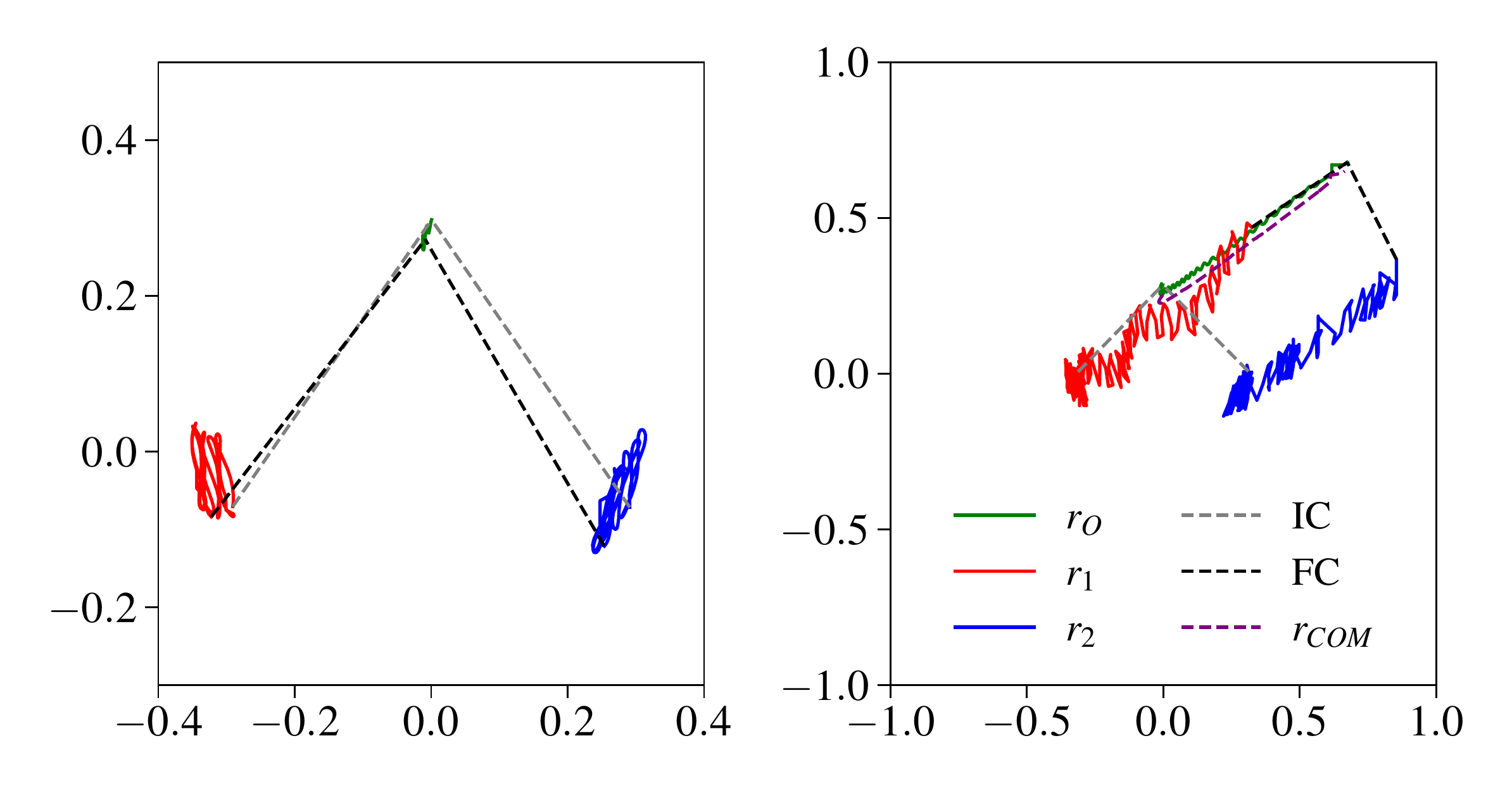}
	\caption{
		Analog water molecule simulation results. The left panel shows a closeup of a small
		excitation around the resting positions. The red and blue trajectories represent the
		two hydrogen atoms. The oxygen atom (green trajectory) is almost at rest. The
		OH binding is indicated by the dashed lines. The grey dashed
		lines show the initial condition (IC), while the dark dashed lines indicate the
		final condition (FC). 
		The right panel shows a long time run which clearly shows the center of mass
		($r_{COM}$, purple line) drifting along with the oxygen $r_O$.
	}\label{fig:results}.
\end{figure}

The model produces reasonable results for small perturbations, as shown in
Figure  \ref{fig:results} for a simulation time of $t_\text{final}=60$ms.

\subsection{Energy and time demand}
The bending patterns have roughly a wall clock time frequency of
$f\sim 100$\,Hz, while the water bending normal modes have a frequency of
$\nu\sim 32$\,Ghz in nature. This gives an effective ratio of about
$k \sim 10^{-9}$ for the analog implementation, or a simulation time per real
clock time ratio of $100/32 \sim 3\,$ns/s or 260\,$\mu$s/day. This number has to
be compared with typical simulations on digital computers, which achieve a
few hundred ns per day
(see for instance a review \cite{Gecht2020} and references therein, such as
\cite{SalomonFerrer2012}).

It is a crucial property of analog computing that this speed factor is invariant of
the size of the equation (and circuitry, respectively)! That it, it is virtually 
independent of both the number of particles simulated as well as the physics
included (\cf Section \ref{sec:charmm}).
In contrast, the power (and thus energy) requirements scale linearly with the size of the circuit
\cite{kht2020}.
The \emph{Analog Paradigm Model-1} power requirement for the present circuit is
measured as $P_A = 21$W.
The overall energy requirement for a single bending wave
is therefore $E_A = P_A T \sim 210$\,mJ.


\begin{figure}
	\includegraphics[width=\linewidth]{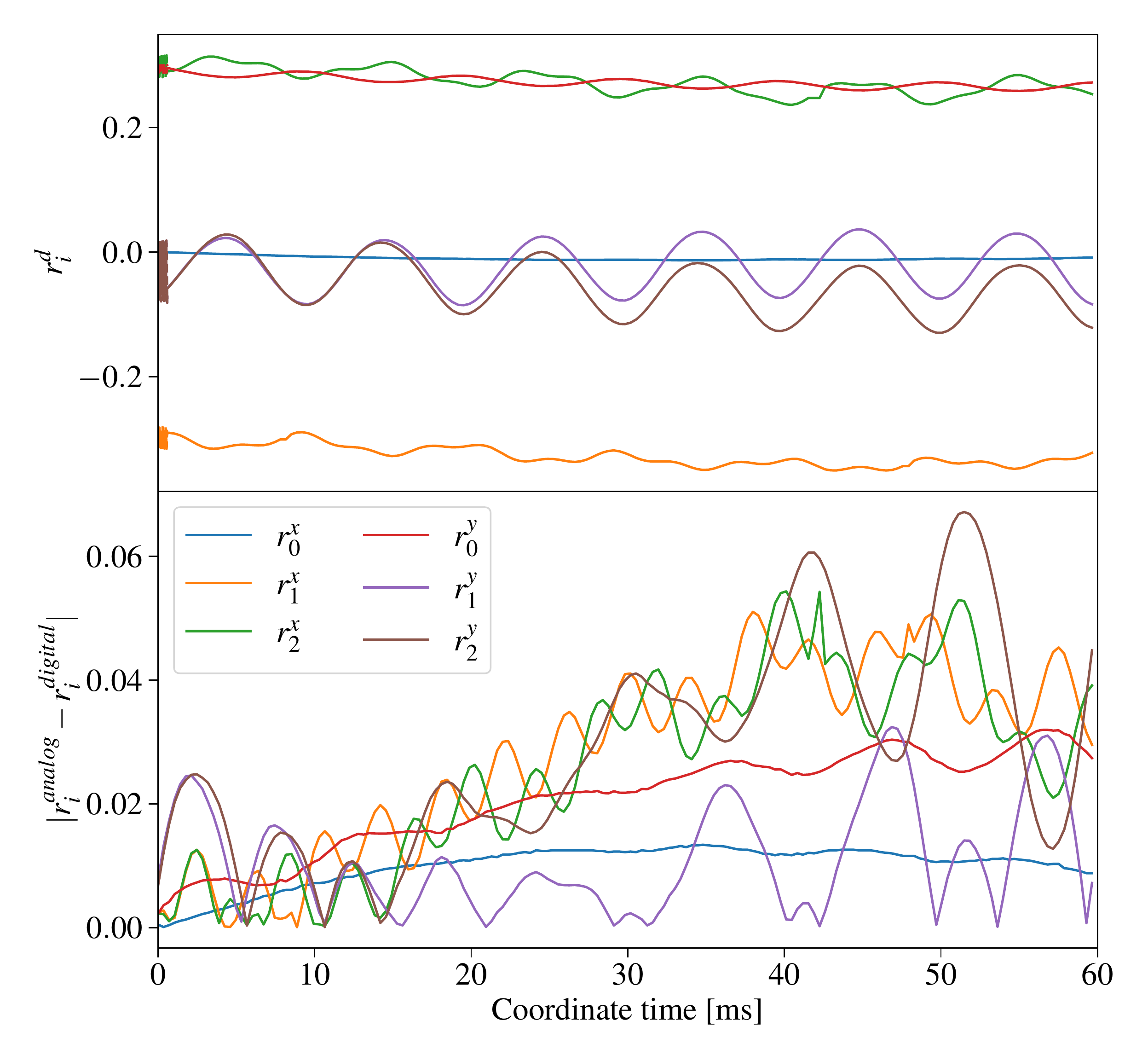}
	\caption{Upper panel: Analog (solid) vs. digital (dashed) time evolution.
		Lower panel: Absolute error (L1 norm) of digital vs. analog resuls for all
        particle coordinates $r^d_i$.
    }\label{fig:analog-digital-comparison}
\end{figure}

\subsection{Upscaling: Towards real-life molecular dynamics}
The overall number of computing elements required for this simulation can be obtained from
the system equation \eqref{eq:analog-equations} or the circuit schematics
(Appendix~\ref{apx:circuit}). It scales with spatial dimensions
$D$, number of atoms $N$ and reduced number of degrees of
freedom $M=(N^2-N)/2$ as
\begin{alignat*}{8}
	(2D+1)M + 2 = 5 \cdot 3 + 2= 17 &\quad \times  \text{SUM} ~(\Sigma) \\
	(2D+2) M = 6 \cdot 3 =  18 &\quad \times \text{MUL} ~(\Pi)  \\
	M = 3 &\quad \times \text{SQRT} ~(\sqrt{\cdot})  \\
	2ND = 2 \cdot 3 \cdot 2 =  12 &\quad \times \text{INT} ~(\smallint)\,.
\end{alignat*}
Ignoring the different types of computing elements, one comes up
with 
$\mathcal{O}(8 N^2)$ computing elements in $D=3$ dimensions. For
instance, a typical water box in $D=3$ with a size of $(\SI{30}{\angstrom})^3$
and $N \sim 10^4$ atoms requires $7\times 10^6$ computing elements.

One can estimate the performance of such a simulation carried out on a future analog
computer on a chip. To do so, the reasoning of \cite{kht2020} is employed. Moving
from an \emph{Analog Paradigm} Model-1 analog computer to a proposed analog 
computer on chip will decrease energy consumption by at least a factor of 40 
with an increase in speed by a factor of at least 150. This results in a 
power demand for a fully integrated analog computer simulating $N\sim10^4$ 
atoms of $P = 10^4/3 \times 20\text{W}/40 \approx 1600$W with a simulation per
real clock time ratio of more than $10^7$\,ns/day!

While this number of computing elements is definetly in the ballpark of
future analog computing chips, the number of interconnections is
certainly the limiting factor. The quadratic dependency on $N$ is in fact dominated by the 
all-to-all force coupling $F_{ij}$ in Newtonian mechanics.
Typically, a long distance cutoff is
introduced in $N$-body simulations involving inverse square law forces
(such as molecular dynamics or galaxy simulations in cosmology).
A similar approach has to be chosen in an analog implementation.
For instance, the authors of \cite{Mark2001} do a spherical cutoff
at $L=\SI{12}{\angstrom}$. Concepts for implementing cutoffs on an
analog computer (where the wiring is ,,hard coded`` at programming time)
are left as an open question for the future.


\subsection{Comparison with numerical time evolution}
A classical algorithmic, \ie numerical, integration scheme for
\eqref{eq:analog-equations} is to evaluate the
time integrals in an iterative manner, for instance with the well-known
primitive Euler scheme:
\begin{equation}
	y(t) = \int f(y(t),t)\,\mathrm d t 
	~\rightsquigarrow~ y_{k+1} = y_k + f(y_k) \Delta t \,.
\end{equation}
Here, $k$ is an index denoting the values at a discretized time $t = k \Delta t$.
The time discretization $\Delta t$ can be basically chosen arbitrarily small, which
is unique to digital computers.

In practice, a higher order (still explicit) Runge-Kutta scheme may be adopted to
integrate both the velocities $\pmb v_i^d$ and subsequently the positions $\pmb r_i^d$ in time. Such an
attempt has an intrinsic inexactness since the underlying ODE \eqref{eq:nbody}
is of second order in time. Spurious damping or excitation of the solutions
(in general an exponential error) will be introduced at any convergence order
\cite{BUTCHER1996247}.
Since classical 3-body problems are known to exhibit chaotic behaviour, long-term
evolutions are inevitably unstable and unsuitable for long time evolutions. This includes
phenomena such as drifting of the center of mass.

In practice, we observe similar stability problems on our analog computer as with traditional
explicit numerical integration schemes. For instance, for longer simulation times
(such as the right panel in Figure~\ref{fig:results}, which shows $t_\text{final}=300$ms),
we see a drift of the center of mass.
This systematic error is reflected in the error plot in Figure \ref{fig:analog-digital-comparison},
which shows a comparision with a high resolution numerical time evolution
($\Delta t = 0.005$ with a first order Euler scheme).

For this long standing problem in computational physics,
methods have been developed to improve accuracy and maintain the validity of the
physical solution. 
A remarkable scheme for Hamiltonian systems such as \eqref{eq:nbody}
are \emph{geometric} or \emph{symplectic} integrators which, among other things, preserve the phase space
volume of the solution and the geometric structure in space-time
\cite{GeometricIntegrators2006}.
The popular symplectic splitting method is basically still explicit. This approach has not
been used in this work due to limitations in size of the available analog computers.
However, in principle approaches like this can be implemented on future analog computing circuits in a
straightforward manner.

\section{Conclusion}\label{sec:conclusion}

We have demonstrated the first analog computer implementation of a simple water model aimed towards
solvent theory and molecular dynamics. In order to implement an actual analog
computer circuit, a simple model was chosen, which implements electrostatic interactions and which
models molecular stretching and bonding in a quadratic potential in Cartesian coordinates.
Since the computer used in this study has just enough computing elements to implement a single water molecule,
intermolecular interactions such as Lennard-Jones potentials were not taken into account.
Special focus was given to the issue how to rigorously scale such a problem for an analog
computer. Thanks to hardware-supported scientific (floating point) number representation on
digital computers, this topic is discussed only rarely in numerical approaches, for instance
in the context of catastrophic cancellation (loss of significance).
Number representation and loss of resolution are problems not yet solved within this work, but
approaches exist to improve these properties without loosing the advantages such as fully
parallel evaluation and exceptionally good energy efficiency.
We also expect that a number of techniques (and experiences) in numerically solving molecular dynamics can
be applied on analog-digital hybrid approaches, such as more robust higher order
integration schemes or implementing long-distance effects (spatial cutoffs).

There  are also completely different approaches, such as the application of
artificial intelligence approaches in molecular dynamics \cite{Abiodun2018}.
For instance, in density functional theory (DFT) approaches, Behler and Parrinello \cite{Behler2007}
introduce a new breed of neural network model of DFT. This gives the energy as a function for all atomic positions in
systems with arbitrary size and of various order of mangitude, which is faster than traditional
DFT approaches. The high level of accuracy of the neural network method is
shown for bulk silicon compared with DFT. It may turn out that implementing analog neural networks
which solve DFT or MD can be a very fruitful approach combining the energy
efficiency and fast time to solution shown by analog computers without the issues of stability, convergence and resolution.

\appendices


\bibliography{AnalogWater}

\begin{thebibliography}{10}
\providecommand{\url}[1]{#1}
\csname url@samestyle\endcsname
\providecommand{\newblock}{\relax}
\providecommand{\bibinfo}[2]{#2}
\providecommand{\BIBentrySTDinterwordspacing}{\spaceskip=0pt\relax}
\providecommand{\BIBentryALTinterwordstretchfactor}{4}
\providecommand{\BIBentryALTinterwordspacing}{\spaceskip=\fontdimen2\font plus
\BIBentryALTinterwordstretchfactor\fontdimen3\font minus
  \fontdimen4\font\relax}
\providecommand{\BIBforeignlanguage}[2]{{%
\expandafter\ifx\csname l@#1\endcsname\relax
\typeout{** WARNING: IEEEtran.bst: No hyphenation pattern has been}%
\typeout{** loaded for the language `#1'. Using the pattern for}%
\typeout{** the default language instead.}%
\else
\language=\csname l@#1\endcsname
\fi
#2}}
\providecommand{\BIBdecl}{\relax}
\BIBdecl

\bibitem{kht2020}
S.~{K{\"o}ppel}, B.~{Ulmann}, L.~{Heimann}, and D.~{Killat}, ``Using analog
  computers in today's largest computational challenges,'' \emph{arXiv
  e-prints}, p. arXiv:2102.07268, Feb. 2021.

\bibitem{Rapaport2004}
\BIBentryALTinterwordspacing
D.~C. Rapaport, \emph{The Art of Molecular Dynamics Simulation}.\hskip 1em plus
  0.5em minus 0.4em\relax Cambridge University Press, Apr. 2004. [Online].
  Available: \url{https://doi.org/10.1017/cbo9780511816581}
\BIBentrySTDinterwordspacing

\bibitem{Guvench2008}
O.~Guvench and J.~MacKerell, ``Comparison of protein force fields for molecular
  dynamics simulations,'' \emph{Methods Mol Biol.}, no. 443, pp. 63--88, 2008.

\bibitem{Lopes2014}
\BIBentryALTinterwordspacing
P.~E.~M. Lopes, O.~Guvench, and A.~D. MacKerell, ``Current status of protein
  force fields for molecular dynamics simulations,'' in \emph{Methods in
  Molecular Biology}.\hskip 1em plus 0.5em minus 0.4em\relax Springer New York,
  Sep. 2014, pp. 47--71. [Online]. Available:
  \url{https://doi.org/10.1007/978-1-4939-1465-4_3}
\BIBentrySTDinterwordspacing

\bibitem{Mackerell2004}
\BIBentryALTinterwordspacing
A.~D. Mackerell, ``Empirical force fields for biological macromolecules:
  Overview and issues,'' \emph{Journal of Computational Chemistry}, vol.~25,
  no.~13, pp. 1584--1604, 2004. [Online]. Available:
  \url{https://doi.org/10.1002/jcc.20082}
\BIBentrySTDinterwordspacing

\bibitem{rahmann71}
\BIBentryALTinterwordspacing
A.~Rahman and F.~H. Stillinger, ``Molecular dynamics study of liquid water,''
  \emph{The Journal of Chemical Physics}, vol.~55, no.~7, pp. 3336--3359, 1971.
  [Online]. Available: \url{https://doi.org/10.1063/1.1676585}
\BIBentrySTDinterwordspacing

\bibitem{hadley12}
K.~Hadley and C.~Mccabe, ``Coarse-grained molecular models of water: A
  review,'' \emph{Molecular simulation}, vol.~38, pp. 671--681, 07 2012.

\bibitem{brooks2009}
\BIBentryALTinterwordspacing
B.~R. Brooks \emph{et~al.}, ``Charmm: The biomolecular simulation program,''
  \emph{Journal of Computational Chemistry}, vol.~30, no.~10, pp. 1545--1614,
  2009. [Online]. Available:
  \url{https://onlinelibrary.wiley.com/doi/abs/10.1002/jcc.21287}
\BIBentrySTDinterwordspacing

\bibitem{brooks1983}
\BIBentryALTinterwordspacing
B.~R. Brooks, R.~E. Bruccoleri, B.~D. Olafson, D.~J. States, S.~Swaminathan,
  and M.~Karplus, ``Charmm: A program for macromolecular energy, minimization,
  and dynamics calculations,'' \emph{Journal of Computational Chemistry},
  vol.~4, no.~2, pp. 187--217, 1983. [Online]. Available:
  \url{https://onlinelibrary.wiley.com/doi/abs/10.1002/jcc.540040211}
\BIBentrySTDinterwordspacing

\bibitem{Joergsen1983}
\BIBentryALTinterwordspacing
W.~L. Jorgensen, J.~Chandrasekhar, J.~D. Madura, R.~W. Impey, and M.~L. Klein,
  ``Comparison of simple potential functions for simulating liquid water,''
  \emph{The Journal of Chemical Physics}, vol.~79, no.~2, pp. 926--935, 1983.
  [Online]. Available: \url{https://doi.org/10.1063/1.445869}
\BIBentrySTDinterwordspacing

\bibitem{ideasOfQuantumChemistryBook}
\BIBentryALTinterwordspacing
``Chapter 7 - motion of nuclei,'' in \emph{Ideas of Quantum Chemistry},
  L.~Piela, Ed.\hskip 1em plus 0.5em minus 0.4em\relax Amsterdam: Elsevier,
  2007, pp. 275--323. [Online]. Available:
  \url{https://www.sciencedirect.com/science/article/pii/B9780444522276500089}
\BIBentrySTDinterwordspacing

\bibitem{Mark2001}
\BIBentryALTinterwordspacing
P.~Mark and L.~Nilsson, ``Structure and dynamics of the {TIP}3p, {SPC}, and
  {SPC}/e water models at 298 k,'' \emph{The Journal of Physical Chemistry A},
  vol. 105, no.~43, pp. 9954--9960, Nov. 2001. [Online]. Available:
  \url{https://doi.org/10.1021/jp003020w}
\BIBentrySTDinterwordspacing

\bibitem{Model1Handbook}
\BIBentryALTinterwordspacing
B.~Ulmann, \emph{Model-1 Analog Computer Handbook/User Manual}, 2019. [Online].
  Available: \url{http://analogparadigm.com/downloads/handbook.pdf}
\BIBentrySTDinterwordspacing

\bibitem{Westphal1995}
\BIBentryALTinterwordspacing
L.~C. Westphal, \emph{BIBO stability and simple tests}.\hskip 1em plus 0.5em
  minus 0.4em\relax Boston, MA: Springer US, 1995, pp. 351--370. [Online].
  Available: \url{https://doi.org/10.1007/978-1-4615-1805-1_14}
\BIBentrySTDinterwordspacing

\bibitem{Sarpeshkar1998}
\BIBentryALTinterwordspacing
R.~Sarpeshkar, ``Analog versus digital: Extrapolating from electronics to
  neurobiology,'' \emph{Neural Computation}, vol.~10, no.~7, pp. 1601--1638,
  Oct. 1998. [Online]. Available:
  \url{https://doi.org/10.1162/089976698300017052}
\BIBentrySTDinterwordspacing

\bibitem{ap2}
B.~Ulmann, \emph{Analog and Hybrid Computer Programming}.\hskip 1em plus 0.5em
  minus 0.4em\relax De Gruyter, 2020.

\bibitem{Gecht2020}
\BIBentryALTinterwordspacing
M.~Gecht, M.~Siggel, M.~Linke, G.~Hummer, and J.~K\"{o}finger, ``{MDBenchmark}:
  A toolkit to optimize the performance of molecular dynamics simulations,''
  \emph{The Journal of Chemical Physics}, vol. 153, no.~14, p. 144105, Oct.
  2020. [Online]. Available: \url{https://doi.org/10.1063/5.0019045}
\BIBentrySTDinterwordspacing

\bibitem{SalomonFerrer2012}
\BIBentryALTinterwordspacing
R.~Salomon-Ferrer, D.~A. Case, and R.~C. Walker, ``An overview of the amber
  biomolecular simulation package,'' \emph{Wiley Interdisciplinary Reviews:
  Computational Molecular Science}, vol.~3, no.~2, pp. 198--210, Sep. 2012.
  [Online]. Available: \url{https://doi.org/10.1002/wcms.1121}
\BIBentrySTDinterwordspacing

\bibitem{BUTCHER1996247}
\BIBentryALTinterwordspacing
J.~Butcher, ``A history of runge-kutta methods,'' \emph{Applied Numerical
  Mathematics}, vol.~20, no.~3, pp. 247--260, 1996. [Online]. Available:
  \url{https://www.sciencedirect.com/science/article/pii/0168927495001085}
\BIBentrySTDinterwordspacing

\bibitem{GeometricIntegrators2006}
\BIBentryALTinterwordspacing
\emph{Geometric Numerical Integration}.\hskip 1em plus 0.5em minus 0.4em\relax
  Springer-Verlag, 2006. [Online]. Available:
  \url{https://doi.org/10.1007/3-540-30666-8}
\BIBentrySTDinterwordspacing

\bibitem{Abiodun2018}
\BIBentryALTinterwordspacing
O.~I. Abiodun, A.~Jantan, A.~E. Omolara, K.~V. Dada, N.~A. Mohamed, and
  H.~Arshad, ``State-of-the-art in artificial neural network applications: A
  survey,'' \emph{Heliyon}, vol.~4, no.~11, p. e00938, Nov. 2018. [Online].
  Available: \url{https://doi.org/10.1016/j.heliyon.2018.e00938}
\BIBentrySTDinterwordspacing

\bibitem{Behler2007}
\BIBentryALTinterwordspacing
J.~Behler and M.~Parrinello, ``Generalized neural-network representation of
  high-dimensional potential-energy surfaces,'' \emph{Physical Review Letters},
  vol.~98, no.~14, Apr. 2007. [Online]. Available:
  \url{https://doi.org/10.1103/physrevlett.98.146401}
\BIBentrySTDinterwordspacing

\end{thebibliography}
\bibliographystyle{IEEEtran}
%
%
%
\begin{IEEEbiography}
	[{\includegraphics[width=1in,height=1.25in,clip,keepaspectratio]{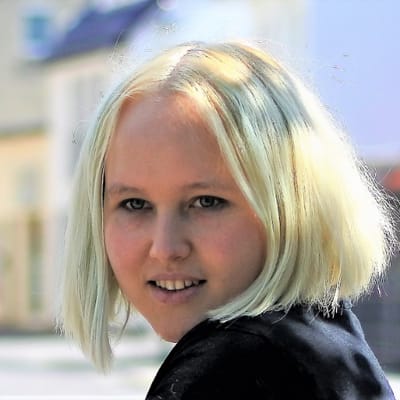}}]
	{Alexandra Krause}
    is a master student in physics at the Physics department of Freie Universität Berlin. Her research interests include computational biophysics, quantum and analog computing.
\end{IEEEbiography}%
\begin{IEEEbiography}
	[{\includegraphics[width=1in,height=1.25in,clip,keepaspectratio]{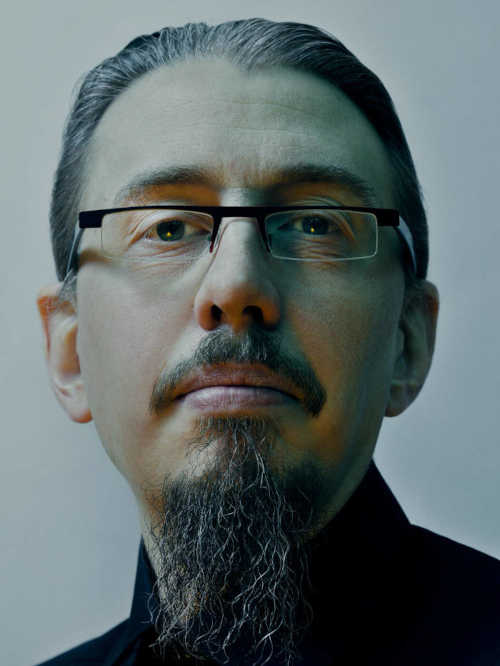}}]
	{Bernd Ulmann}
	is professor at the FOM University of Applied Sciences for Economics and Management. He also is guest professor and lecturer at the Institute of Medical Systems Biology at Ulm University. His main area of interest is analog and hybrid computing.
\end{IEEEbiography}%
\begin{IEEEbiography}
	[{\includegraphics[width=1in,height=1.25in,clip,keepaspectratio]{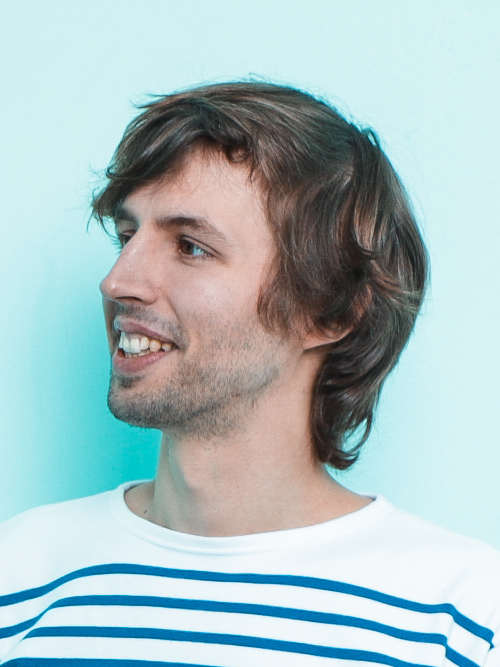}}]
	{Sven K\"oppel}
	is a theoretical high energy physicist and Chief Scientific Officer at
	{anabrid GmbH}, Berlin. His main area of expertise is computational
	physics and high performance computing. He graduated at Goethe University,
	Frankfurt, on general relativistic simulations of black holes, neutron
	stars and gravitational waves.	
\end{IEEEbiography}%
\end{document}